\documentclass[pdflatex,sn-basic]{sn-jnl}

\usepackage{graphicx}%
\usepackage{multirow}%
\usepackage{amsmath,amssymb,amsfonts}%
\usepackage{amsthm}%
\usepackage{mathrsfs}%
\usepackage[title]{appendix}%
\usepackage{xcolor}%
\usepackage{textcomp}%
\usepackage{manyfoot}%
\usepackage{booktabs}%
\usepackage{algorithm}%
\usepackage{algorithmicx}%
\usepackage{algpseudocode}%
\usepackage{listings}%
\usepackage{mdframed}
\usepackage{array}
\usepackage{verbatim}

\theoremstyle{thmstyleone}%
\theoremstyle{thmstyletwo}%

\theoremstyle{thmstylethree}%

\begin{document}

\title[Article Title]{The Semiotic Channel Principle: Measuring the Capacity for Meaning in LLM Communication}


\author[1]{\fnm{Davide} \sur{Picca}}\email{davide.picca@unil.ch}



\affil[1]{\orgdiv{} \orgname{University of Lausanne}, \orgaddress{\street{} \city{Lausanne}, \postcode{} \state{} \country{Switzerland}}}



\abstract{
In this paper, we propose a novel framework based on semiotic principles for analyzing Large Language Models (LLMs). We propose conceptualizing them not as artificial minds, but as stochastic semiotic engines whose outputs require active, asymmetric interpretation by human users. We formalize the dynamic relationship between expressive richness and interpretive stability using information-theoretic tools: \emph{semiotic breadth} is quantified as the source entropy, while \emph{decipherability} is measured by the mutual information between messages and their human interpretations within a given audience and context. Crucially, we introduce a generative complexity parameter ($\lambda$) that tunes the message distribution, making both breadth and decipherability functions of $\lambda$. The fundamental trade-off between these dimensions is thus modeled as an \emph{emergent property} of their distinct functional responses to $\lambda$. We introduce the notion of a \emph{semiotic channel}, parameterized by audience and context, and posit a capacity constraint on meaning transmission. This capacity is operationally defined as the maximum decipherability achievable by optimizing $\lambda$. This operational reframing shifts analysis from opaque model internals to observable textual artifacts, enabling empirical measurement of breadth (via entropy estimates) and decipherability (via interpretation distributions and conditional entropy).  We demonstrate the utility of this framework through four applications: (i) semiotic profiling and certification of models; (ii) prompt and context design for optimizing interpretive outcomes; (iii) risk analysis and disinformation detection based on structural ambiguity; and (iv) adaptive systems that dynamically adjust their semiotic profile to audience and task. We conclude by discussing methodological and theoretical limitations, and argue that a capacity-based semiotic approach provides a rigorous and actionable toolkit for understanding, evaluating, and designing LLM-mediated communication.
}




\maketitle

\section{Introduction}

Large Language Models (LLMs) have rapidly transformed language technologies, blurring the boundary between ``human text'' and ``artificial text'' \citep{bender2021, mitchell2023}. Their ability to generate coherent, creative, and contextually appropriate outputs raises fundamental questions about meaning, interpretation, and the nature of communication. When a text is produced without authorial intention or embodied experience, how should we understand its status as a sign? What new interpretive challenges arise for human users engaging with such artifacts?

While computational, ethical, and cognitive perspectives on LLMs abound, the semiotic dimension---the science of signs and signification \citep{eco1976, peirce1931, saussure1916}---remains underexplored. Most current debates focus on the internal mechanisms or ethical risks of LLMs, often neglecting the fundamental processes of meaning-making that occur at the interface between artificial outputs and human interpreters. Classical arguments, such as Searle's ``Chinese Room'' \citep{searle1980}, highlight the gap between syntax and semantics, but a systematic framework for analyzing LLM outputs as semiotic artifacts is still lacking.

This paper addresses that gap by advancing a semiotic perspective on LLMs. We propose that every LLM output functions as a sign, and that every human–machine interaction constitutes an asymmetric ``semiotic game'' \citep{eco1979}, in which the machine produces signifiers based on statistical correlations, while the human user is responsible for closing the circuit of meaning \citep{peirce1931}. LLMs, in this view, are stochastic semiotic engines: they manipulate linguistic forms without genuine understanding, and their outputs require active interpretive cooperation from users \citep{picca2025}. This dynamic foregrounds the need to revisit classic semiotic categories, such as ``polysemy,'' ``open text,'' and ``interpretive cooperation'' \citep{barthes1970, eco1979}, to develop new tools for analyzing the emergent properties of artificial texts.

A central challenge in this context is the dynamic relationship between expressive richness and interpretive stability. Drawing on Floridi's work on the trade-off between certainty and purpose in information systems \citep{floridi2024}, we transpose this dialectic to the semiotic domain: the tension between a text's expressive richness and its interpretive stability. LLM-generated texts can be highly diverse and creative, but this very richness often comes at the cost of increased ambiguity and interpretive dispersion. Conversely, maximally unambiguous texts may sacrifice expressive power. To formalize this dynamic, we introduce a generative complexity parameter ($\lambda$) that controls the message distribution. 


Analogously to communication theory, where a channel represents the medium through which messages are transmitted from a source to a receiver, in our semiotic framework, the audience ($\mathcal{A}$) and a context ($\mathrm{Ctx}$) together form a \emph{semiotic channel} that maps model-generated messages to human interpretations ($Int$). Just as traditional communication channels have a capacity limit---a maximum rate at which information can be reliably transmitted---the semiotic channel also possesses a fundamental capacity constraint. This capacity $\mathcal{C}$ is operationally defined as the maximum decipherability achievable by optimizing the generative complexity $\lambda$ for a given channel. Consequently, in the context of LLMs, even the most precisely engineered prompts cannot guarantee unlimited interpretive precision when the audience-context channel imposes inherent limitations on meaning transmission.

This framework externalizes analysis from the opaque internal processes of LLMs to their observable outputs, providing an operational basis for measurement: semiotic breadth via entropy-rate proxies, decipherability via empirical distributions of human interpretations, and ambiguity via conditional entropy. 
By grounding our approach in both semiotic theory and information-theoretic formalism, we offer a coherent and testable toolkit for understanding, evaluating, and designing communicative ecosystems mediated by LLMs. The remainder of this paper develops this framework in detail, explores its practical applications---from model evaluation and prompt engineering to interpretive risk analysis and adaptive system design---and discusses its methodological and theoretical limitations.

\section{Theoretical Foundations: Why Semiotics for LLMs?}

A rigorous analysis of LLM outputs requires a shift in perspective: rather than interrogating the ``intelligence'' or ``consciousness'' of these systems, we must focus on their operational function as generators of signs. In this view, an LLM is not a ``thinking'' entity, but a \emph{stochastic semiotic engine} \citep{picca2025}. Its outputs are not guided by intention or world understanding, but are the result of probabilistic sampling from distributions learned over vast corpora. Each token is selected based on statistical correlations, producing text that is syntactically and stylistically plausible, yet fundamentally devoid of intentionality or embodied meaning \citep{bender2021, mitchell2023}.

This distinction is crucial: for an LLM, the concept of ``cat'' is not grounded in sensory experience, but encoded as a high-dimensional vector defined by its statistical proximity to other words. In contrast, a human's concept of a cat is shaped by a lifetime of embodied, multisensory interactions. The LLM's ``knowledge'' is thus a map of correlations---a statistical model of language without access to the territory of lived experience.

This radical dissociation between signifier and referent makes classic semiotic models, such as Peirce's semiotic triangle \citep{peirce1931}, especially relevant for analyzing human–machine communication. In human discourse, meaning arises from the interplay between the sign (the word), the object (the real or fictional-world referent), and the interpretant (the concept in the mind of the listener), all grounded in shared, embodied experience \citep{picca2020}. For LLMs, this triad is incomplete: the model excels at producing the sign, but lacks direct access to the object and cannot form an independent interpretant. This structural gap explains phenomena such as ``hallucinations,'' where the model generates plausible but unverifiable statements---not as errors, but as natural consequences of its design and lack of grounding \citep{bender2021, picca2024}.

Meaning-making in LLM-mediated communication is therefore fundamentally \emph{asymmetric}. The machine produces signs ``orphaned'' of both experiential grounding and intentionality; it is the human user who must close the circuit of semiosis, drawing on their own knowledge, context, and interpretive frameworks to construct meaning. This process is not a dialogue between two minds, but an act of \emph{asymmetric interpretive cooperation}, in which the user actively fills semantic gaps and may even project intentionality onto the system---a tendency known as the ``ELIZA effect'' \citep{hofstadter1995}.

This dynamic resonates with Eco's concepts of the ``open work'' and the ``Lector in Fabula'' \citep{eco1979, eco1989-openwork}. LLMs, by their probabilistic nature, generate intrinsically ``open'' texts: each output is a single realisation from a vast space of possible continuations, and is susceptible to a multiplicity of readings. Unlike the deliberate ambiguity of avant-garde literature, however, the openness of LLM-generated text is non-intentional and unbounded, limited only by the statistical properties of the training data. In human-authored works, openness is ultimately constrained by the author's project and cultural context; in LLMs, it is a structural property of the model.

The user's role, then, is analogous to Eco's model reader: to navigate this openness, actualize a specific meaning, and impose coherence on the output. In this sense, prompt engineering becomes a form of semiotic design---the art of constructing a context ($\mathrm{Ctx}$) that prunes the tree of interpretive possibilities and guides the model toward producing useful and relevant signs \citep{picca2025}.

Within this framework, we define two key operational dimensions:
\begin{itemize}
    \item \emph{Semiotic breadth} ($S$): the potential richness and diversity of meanings a text can evoke, formalized as the entropy of the message source.
    \item \emph{Decipherability} ($D$): the degree to which a text, in a given context and for a given audience, guides interpreters toward a convergent understanding, formalized as the mutual information between messages and interpretations.
\end{itemize}

Understanding LLMs as semiotic machines thus means formalizing the rules of this new interpretive game. It requires moving from intuition to measurement, and developing tools to predict, evaluate, and optimize the collaborative creation of meaning that emerges between a non-conscious sign producer and a fully human interpreter.

\section{A Semiotic Capacity Framework: From Theory to Formalization}

LLM-mediated communication is fundamentally a process of asymmetric interpretive cooperation, shaped by the inherently ``open'' nature of model outputs. To move from intuition to rigorous analysis, we must formalize the dynamics that govern meaning-making in this setting. Drawing on Floridi’s work on the trade-off between certainty and purpose in information systems \citep{floridi2024}, we transpose this dialectic to the semiotic domain: the tension between a text’s expressive richness and its interpretive stability.

Our central claim is that this trade-off can be precisely characterized using information-theoretic tools. Rather than focusing on the opaque internal mechanisms of LLMs, we analyze their outputs as observable semiotic artifacts, governed by measurable communicative constraints. This reframing shifts the object of study from the ``mind-like machine'' to the ``meaning-making artifact,'' enabling the development of externally-focused metrics for evaluating and designing communicative systems.

\subsection{Formalizing the Semiotic Dynamic}

To move from intuition to rigorous analysis, we must formalize the dynamics that govern meaning-making. We do so by defining two key operational dimensions and a control parameter that governs their relationship.

\paragraph{Semiotic Breadth ($S$).}
This quantifies the richness, diversity, and complexity of a text. Formally, semiotic breadth $S$ is defined as the entropy of the generated message $M$, for a given audience $\mathcal{A}$ and context $\mathrm{Ctx}$:
\[
S := H_{\mathcal{A},\mathrm{Ctx}}(M)
\]
High entropy indicates a source capable of generating a wide variety of diverse and unpredictable messages, reflecting greater expressive richness.

\paragraph{Semiotic Decipherability ($D$).}
This measures the degree to which a text leads to a stable and shared interpretation. Formally, decipherability $D$ is the mutual information between messages $M$ and their interpretations $Int$:
\[
D_{\mathcal{A},\mathrm{Ctx}} := I_{\mathcal{A}, Ctx}(M;Int) = H_{\mathcal{A},\mathrm{Ctx}}(Int) - H_{\mathcal{A},\mathrm{Ctx}}(Int \mid M)
\]
It quantifies the reduction in uncertainty about an interpretation that results from observing a message.

\begin{figure}
    \centering
    \includegraphics[width=1\linewidth]{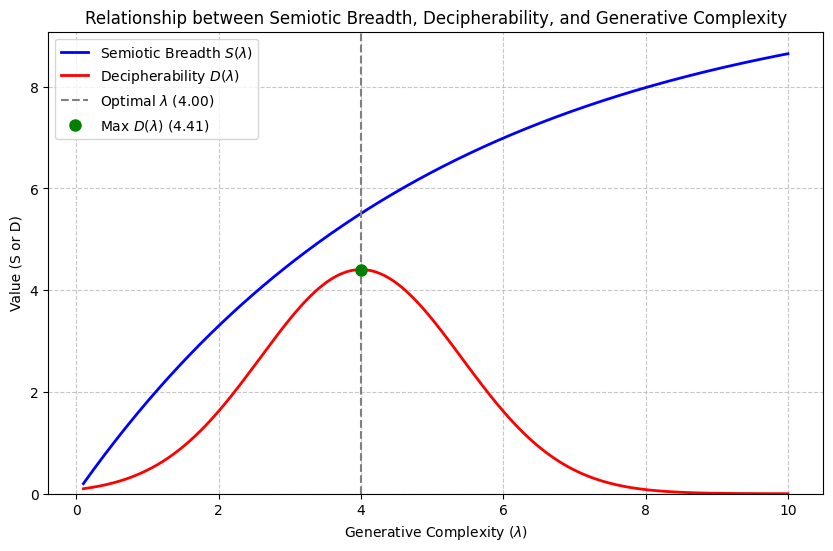}
    \caption{Explanatory example of the relationship between Semiotic Breadth, Decipherability, and Generative Complexity}
    \label{fig:SD}
\end{figure}

\paragraph{Generative Complexity ($\lambda$) and the Emergent Trade-off.}
The relationship between breadth and decipherability is not static. We introduce a control parameter $\lambda$ representing the \emph{generative complexity} of the source (e.g., LLM sampling temperature). This parameter tunes the message distribution $P_\lambda(M)$, making both $S$ and $D$ functions of $\lambda$. We postulate two hypotheses about these functions and depicted in Fig. \ref{fig:SD}:
\begin{enumerate}
    \item Semiotic breadth $S(\lambda)$ is a monotonically increasing function of complexity.
    \item Semiotic decipherability $D(\lambda)$ is the decipherability evaluated under the source distribution $P_{\lambda}(M)$ , which is unimodal as a function of the generative complexity $\lambda$, achieving a maximum at an optimal level $\lambda_{\mathrm{opt}}$ for a given channel $(\mathcal{A}, \mathrm{Ctx})$.
\end{enumerate}

The trade-off emerges directly from this model. To maximize richness ($S$), one must increase $\lambda$. However, rising $\lambda$ beyond $\lambda_{\mathrm{opt}}$ causes decipherability ($D$) to decrease. This dynamic allows us to refine the concept of channel capacity operationally.

It is crucial to note that this framework primarily evaluates the \emph{communicative fidelity} and \emph{interpretive stability} of LLM outputs within a given semiotic channel. It does not, by itself, assess the \emph{factual correctness}, \emph{truthfulness}, or \emph{semantic validity} of the content generated. A message can be highly decipherable yet factually incorrect, and vice-versa. The framework focuses on the structural properties of meaning transmission, not on the correspondence of signs to an external reality.

\paragraph{Principle 1: Capacity is the Optimal Decipherability.}
The capacity $\mathcal{C}$ of a semiotic channel $(\mathcal{A}, \mathrm{Ctx})$ is the maximum achievable decipherability obtained by optimizing the generative complexity:
\[
\mathcal{C}_{\mathcal{A},\mathrm{Ctx}} := \max_{\lambda} D_{\mathcal{A},\mathrm{Ctx}}(\lambda) = D_{\mathcal{A},\mathrm{Ctx}}(\lambda_{\mathrm{opt}})
\]
The fundamental constraint $D(\lambda) \le S(\lambda)$ holds for all values of $\lambda$.

\paragraph{Principle 2: Residual Ambiguity Constrains Interpretive Coherence (and Error in Specific Contexts).}

The residual uncertainty within a semiotic channel is captured by the conditional entropy $H_{\mathcal{A},\mathrm{Ctx}}(Int \mid M)$, which represents the variability of interpretations that persists even after observing the message $M$. This quantity formalizes interpretive dispersion or the effective polysemy of a text for a given audience and context.

\section{Practical Applications and Implications}

Quantifying semiotic breadth ($S$) and decipherability ($D$) enables a range of concrete applications to enhance the effectiveness, safety, and reliability of LLM-mediated communication. Here, we outline four key areas where this framework informs engineering, governance, and human–AI interaction. Our framework introduces a new dimension: the \emph{semiotic profile} of a model, characterized by its typical $(S, D)$ values across different contexts.

\subsection{Model Evaluation and Semiotic Certification}

Traditional LLM evaluation metrics (e.g., BLEU, ROUGE, MMLU) focus on factual accuracy or similarity to reference texts, and are best suited for closed, well-defined tasks. However, most real-world interactions with LLMs occur in open contexts, where interpretive potential and communicative style are as important as correctness.

By empirically measuring $S(\lambda)$ and $D(\lambda)$ across a range of $\lambda$ values for a given model within a specific semiotic channel ($\mathcal{A}, \mathrm{Ctx}$), we can construct a \emph{semiotic profile}. This profile is not a static $(S, D)$ pair, but rather a characterization of the \emph{functional relationship} between generative complexity and the emergent breadth and decipherability. It describes not just whether a model is ``good'' or ``bad,'' but \emph{how it communicates} across its operational spectrum. For example:
\begin{itemize}
    \item A ``Poet'' model might exhibit a rapidly increasing $S(\lambda)$ curve and a $D(\lambda)$ curve that peaks at a relatively high $\lambda$ but then drops sharply, indicating a preference for richness over stability at higher complexities.
    \item A ``Technician'' model may show a flatter $S(\lambda)$ curve and a $D(\lambda)$ curve that peaks at a lower $\lambda$ and remains high across a broader range of $\lambda$ values, ideal for technical documentation where precision is paramount.
    \item A ``Tightrope Walker'' model would demonstrate $S(\lambda)$ and $D(\lambda)$ curves that are carefully balanced, allowing for a flexible optimization of the trade-off across different tasks.
\end{itemize}

Such profiles enable \emph{semiotic certification}: models can be accompanied by visualizations (e.g., plots of $S(\lambda)$ and $D(\lambda)$ curves, or a "semiotic operating map") of their typical performance in standardized semiotic channels (e.g., legal, medical, creative, conversational). Organizations can then specify minimum decipherability thresholds (e.g., $D(\lambda_{\mathrm{opt}}) > 0.8$) and desired breadth characteristics (e.g., $S(\lambda_{\mathrm{opt}})$ within a certain range) for deployment in sensitive applications, reducing the risk of catastrophic misunderstandings. This allows for a more nuanced selection of models based on their communicative characteristics rather than just their factual accuracy.

\subsection{Prompt Engineering and Context Design}

Our framework transforms prompt engineering from an empirical art into a science of contextual control. Given that $S$ and $D$ are shaped by both model and context, prompt design becomes the primary lever for modulating interpretive outcomes.

Effective prompts act as constraints, reducing ambiguity and guiding the model toward desired communicative profiles. For instance:
- A vague prompt (``Write a marketing email for new software.'') leaves high openness, resulting in high $S$ and potentially low $D$.
- A highly specific prompt (``As a B2B marketing expert, write a 150-word email for the launch of  `SyncFlow'... Use a professional but energetic tone and include a clear call-to-action.'') narrows the interpretive space, increasing $D$.

The framework allows for quantitative assessment of prompt strategies: by measuring changes in $D$ and $S$ as a function of prompt specificity, persona cues, or negative instructions, prompt engineers can systematically optimize for interpretive stability or creative breadth as required by the task and audience.

\subsection{Interpretive Risk Analysis and Disinformation Detection}

The semiotic capacity framework provides a quantitative basis for assessing communicative risk. Texts with high $S$ and low $D$ are structurally ambiguous, making them fertile ground for misinterpretation, disinformation, or viral controversy. Such texts function as ``semiotic Rorschach tests,'' inviting projection and divergent readings.

For example, an LLM-generated statement on a sensitive topic may be interpreted in radically different ways by different subgroups, especially in polarized or heterogeneous audiences. By computing an \emph{interpretive risk score}---for instance, the ratio $S/D$---platforms and regulators can flag content with a high ``attack surface'' for malicious or unintended interpretation. Unlike keyword-based filters, this approach targets the structural properties of the message, making it more robust to circumvention.

\subsection{Adaptive Human–Machine Communication Systems}

The most advanced application of this framework is in the design of adaptive dialogue systems. By continuously estimating $S$ and $D$ in real time, a conversational agent can dynamically adjust its communicative profile to match user needs and context.

For example, the system can:
- Infer audience expertise (e.g., via lexical analysis) and adjust $S$ accordingly, providing more complex responses to experts and simpler, more decipherable outputs to novices.
- Detect interpretive failures (e.g., user requests for clarification) and respond by reducing $S$ and increasing $D$ in subsequent turns.
- Operate in ``safe by default'' mode in high-risk contexts (e.g., medical advice), prioritizing high $D$ and actively seeking user confirmation.

This adaptive feedback loop enables a new generation of conversational AI that is not only fluent, but also communicatively competent---capable of optimizing the trade-off between expressive richness and interpretive stability in response to real-world demands.

Every formal model is a simplification of reality, and its value depends not only on explanatory power but also on a clear understanding of its boundaries. The semiotic capacity framework presented here offers a novel, quantitative approach to LLM-mediated communication, but it is essential to critically assess both its strengths and its limitations.

\section{Methodological Limitations}

Despite its strengths, the framework faces several methodological challenges, particularly in measurement and empirical implementation.

\paragraph{Operationalizing and Measuring Interpretations ($Int$).}
The most significant methodological challenge lies in the rigorous operationalization and measurement of human interpretations ($Int$). Our framework relies on $Int$ as a random variable, whose probability distribution $P(Int)$ and conditional distribution $P(Int \mid M)$ are essential for calculating decipherability ($D$). However, human interpretation is inherently subjective, context-dependent, and difficult to quantify. While methods like crowd-sourcing, qualitative coding, or semantic similarity metrics can provide proxies for interpretive convergence or divergence, these approaches face several issues:
\begin{itemize}
    \item \emph{Granularity}: Defining the ``units" of interpretation (e.g., keywords, sentiment, abstract concepts) and their relationships is complex.
    \item \emph{Bias}: Human annotators bring their own biases and backgrounds, potentially influencing the observed distribution of $Int$.
    \item \emph{Scalability}: Collecting and analyzing large datasets of human interpretations for diverse messages and contexts is resource-intensive.
    \item \emph{Validity}: Ensuring that the chosen operationalization of $Int$ truly captures the theoretical construct of ``interpretation" remains a non-trivial task.
\end{itemize}
The empirical validation of our framework, therefore, hinges critically on the development of robust, scalable, and theoretically grounded methodologies for measuring $Int$.

\paragraph{Proxy-based metrics.}
Beyond the fundamental challenge of $Int$, quantifying other complex constructs such as semiotic breadth necessarily relies on proxies (e.g., lexical diversity, polysemy indices, or automated metaphor detection). These measures, while useful, are approximations and may not capture the full nuance of human interpretation. For example, automated methods often struggle with context-dependent or subtle interpretive phenomena \citep{egloff2019}.


\subsection{Theoretical Limitations}

The framework also rests on theoretical assumptions that delimit its scope.

\emph{Optimization is not always the goal:}  
The model assumes that communicative effectiveness is achieved by optimizing the $S$–$D$ trade-off near the capacity frontier. While appropriate for many pragmatic contexts (e.g., legal, technical, instructional), this assumption does not hold in domains where ambiguity, multiplicity, or interpretive openness are valued---such as art, poetry, or philosophical discourse. In these cases, low decipherability or high ambiguity may be intentional and desirable.

\emph{Static interpretation model:}  
The current formalism treats interpretation as a single, context-fixed event, measuring convergence or dispersion at a given moment. In reality, semiosis is dynamic: meanings evolve as communities develop new interpretive codes or as contexts shift. The present model does not capture this diachronic evolution.

Recognizing these limitations clarifies the intended role of the framework: not as a universal theory of artificial communication, but as a targeted and powerful tool for analyzing and engineering clarity and complexity in contexts where these dimensions are critical.

\section{Conclusion and Future Developments}

In this work, we introduced and formalized a semiotic capacity framework for analyzing and evaluating outputs generated by Large Language Models (LLMs). Moving beyond purely technical or ethical analyses, we foregrounded the fundamental level of signification, conceptualizing LLMs as stochastic semiotic engines whose outputs require active interpretive cooperation from human users.

Our main contribution rests on four strengths. First, \emph{theoretical grounding}: we bridged classical semiotics and generative AI, grounding our analysis in the traditions of Peirce and Eco to argue that meaning in LLM-mediated communication emerges from an asymmetric human–machine semiotic game. Second, \emph{rigorous formalization}: we translated this intuition into a quantifiable model using information theory, transforming the qualitative trade-off between expressive richness (semiotic breadth, $S$) and interpretive stability (decipherability, $D$) into a measurable, falsifiable claim constrained by semiotic channel capacity.

Third, \emph{operationalizability}: the approach is not merely abstract. We provided concrete measurement protocols for estimating $S$, $D$, and channel capacity $\mathcal{C}$. Fourth, \emph{generality and flexibility}: we demonstrated practical utility across domains—model profiling and semiotic certification, systematic prompt and context design, interpretive risk analysis, and the development of adaptive conversational systems capable of dynamically modulating their semiotic profile.

We acknowledge limitations, including reliance on proxy metrics, the need for broad empirical validation across heterogeneous communicative contexts, and the challenges of operationalizing capacity in dynamic, non-stationary environments. Nonetheless, we view this framework as a significant step toward a more rigorous dialogue between the humanities and computer science.

\section{Acknowledgements}

We wish to express our sincere gratitude to Lorenzo Zangari, from the University of Lausanne. His keen eye was invaluable in correcting and refining the mathematical formulation presented here. Beyond this, our fruitful discussions provided stimulating insights that helped shape the core arguments of this paper.

\bibliography{biblio}
\end{document}